\documentclass[12pt, draftclsnofoot, onecolumn]{IEEEtran}
\usepackage{amsmath,amssymb,amsfonts}
\usepackage{amsthm,ulem}
\usepackage{algorithmic}
\usepackage{algorithm}
\usepackage{array}
\usepackage{textcomp}
\usepackage{stfloats}
\usepackage{url}
\usepackage{verbatim}
\usepackage{graphicx}
\usepackage{subcaption}
\usepackage{caption}
\usepackage{cite}
\usepackage[bookmarks=false]{hyperref}
\newtheorem{lemma}{Lemma}
\newtheorem{remark}{Remark}

\newtheorem{definition}{Definition}
\newtheorem{theorem}{Theorem}
\usepackage{xcolor}

\DeclareMathOperator*{\argmax}{arg\,max}

\newcommand{\V}{\mathcal{V}}
\newcommand{\E}{\mathcal{E}}
\newcommand{\K}{\mathcal{K}}

\newcommand{\pnext}{.\textit{nx}}
\newcommand{\pprev}{.\textit{pr}}

\input{widebar_hack.tex}
\begin{document}
\begin{NoHyper}
\title{Proactive Resilient Transmission and Scheduling Mechanisms for mmWave Networks}

\author{\IEEEauthorblockN{Mine Gokce Dogan,~\IEEEmembership{Student Member,~IEEE,} Martina Cardone,~\IEEEmembership{Senior Member,~IEEE,} and Christina Fragouli,~\IEEEmembership{Fellow,~IEEE}}
\thanks{M. G. Dogan and C. Fragouli are with the Electrical and Computer Engineering Department at the University of California, Los Angeles, CA 90095 USA (e-mail: \{minedogan96, christina.fragouli\}@ucla.edu). The research carried out at UCLA was supported in part by the Army Research Laboratory under Co-Operative Agreement \mbox{W911NF-17-2-0196} and by the U.S. National Science Foundation (NSF) awards \mbox{442521-FC-22071} and \mbox{442521-FC-21454}.
M. Cardone is with the Electrical and Computer Engineering Department of the University of Minnesota, MN 55404 USA (e-mail: cardo089@umn.edu). The work of M. Cardone was supported in part by the NSF under Grants \mbox{CCF-2045237} and \mbox{CNS-2146838}.
Part of this work was presented at the 2021 IEEE Military Communications Conference~\cite{MineMilcom} and at the 2022 IEEE International Symposium on Information Theory~\cite{MineISIT22}.
}
}

\maketitle

\begin{abstract}
This paper aims to develop resilient transmission mechanisms to suitably distribute traffic across multiple paths in an arbitrary millimeter-wave (mmWave) network. The main contributions include: (a) the development of \textit{proactive} transmission mechanisms that build resilience against network disruptions in advance, while achieving a high end-to-end packet rate; 
(b) the design of a heuristic path selection algorithm that efficiently selects (in polynomial time in the network size) multiple proactively resilient paths with high packet rates;
and (c) the development of a \textit{hybrid} scheduling algorithm that combines the proposed path selection algorithm with a deep reinforcement learning (DRL) based online approach for decentralized adaptation to blocked links and failed paths.
To achieve resilience to link failures, a state-of-the-art Soft Actor-Critic DRL algorithm, which adapts the information flow through the network, is investigated. The proposed scheduling algorithm robustly adapts to link failures over different topologies, channel and blockage realizations while offering a superior performance to
alternative algorithms.
\end{abstract}
\begin{IEEEkeywords}
Millimeter-wave networks, network capacity approximations,  resilient transmission mechanisms, path selection, deep reinforcement learning, decentralized adaptation.
\end{IEEEkeywords} 
\section{Introduction}
Millimeter Wave (mmWave) (and beyond) is an enabling technology that is playing an increasingly important role in our wireless infrastructure by expanding the available spectrum and enabling multi-gigabit services~\cite{5GMF-16,Heath16,Rangan}. A number of use cases are currently built around multi-hop mmWave networks, such as Facebook's Terragraph network~\cite{Facebook} that uses flexible mmWave backbones to connect clusters of base stations. Other example scenarios
include private networks, such as in shopping centers, airports and enterprises; mmWave mesh networks that use mmWave links as backhaul in dense urban scenarios; military applications employing mobile hot spots; and mmWave based vehicle-to-everything (V2X) services, such as cooperative perception~\cite{Qualcomm-EU,Hur-13,Woo-16}.
Despite the promising aspects of mmWave communication, mmWave links are highly sensitive to blockage, channels may abruptly change and paths get disrupted~\cite{Jain,Samuylov16,Gapeyenko17,Wang17,Raghavan19}. 
It becomes therefore of fundamental importance to deploy transmission mechanisms that are \textit{resilient} against such disruptions.

In this paper, we aim to develop resilient communication mechanisms to suitably distribute the traffic across multiple paths in a mmWave network by building on the so-called 1-2-1 network model that offers a simple, yet informative, model for mmWave networks~\cite{ezzeldin}. 
We first develop \textit{proactive} transmission mechanisms that build resilience against link failures in advance, with the only knowledge of the probability of link failures, while achieving a high end-to-end packet rate (fraction of packets delivered\footnote{We note that coupling our approach with erasure correcting codes allows to translate packet rates to information throughput.}). The probability of link failures can be highly asymmetric in mmWave networks, yet known in advance through accurate models~\cite{Jain,Samuylov16,Gapeyenko17,Wang17,Raghavan19}, which can be leveraged to achieve a superior performance\footnote{The link blockage probability depends on explicitly known factors such as physical distances, and indoor or outdoor propagation.}. Our goal is to identify which paths to use and how to suitably schedule them, so that the packet rate is as large as possible in the presence of link blockages. A challenging aspect is beam scheduling in mmWave networks, where nodes communicate with each other by using beamforming and scheduling. Which nodes should communicate and for how long, is a non-trivial optimization problem. Then, we leverage the proposed proactive transmission mechanisms to develop a {\em hybrid scheduling algorithm} that combines the proactive mechanisms with a deep reinforcement learning (DRL) based online approach for decentralized adaptation to blocked links and failed paths. In particular, we ask to find which paths the source should use and at which rates to connect with the destination. Towards this end, we use a state-of-the-art DRL algorithm called Soft Actor-Critic (SAC) algorithm~\cite{haarnoja2018soft}, to support a desired packet rate between a source and a destination over an arbitrary mmWave network. Our scheduling algorithm does not require knowledge of the network topology or link capacities frequently, thus it is well suited to volatile environments where channel and topology knowledge can be prohibitively costly to acquire.

\subsection{Contributions}
Our main contributions are summarized as follows,
\begin{enumerate}
\item For an arbitrary mmWave network, we characterize both the optimal \mbox{worst-case} and optimal average packet rates
through Linear Programs (LPs), and present an example of resilience-optimal packet rate \mbox{trade-off} curves. For the \mbox{worst-case} optimal packet rate, our method is polynomial in the network size, but exponential in the maximum number of blocked links.
\item We analyze the structure of the LPs and show that, out of an exponential number of paths (in the number of relay nodes $N$) that potentially connect a source to a destination, in 1-2-1 networks we only need to utilize at most $2N + 2$ paths to characterize the optimal average packet rate. We also show that, when the link capacities are all \textit{equal}, the optimal \mbox{worst-case} and the optimal average packet rates can always be achieved by activating edge-disjoint paths. However, when the link capacities are \textit{unequal}, there exist network topologies for which the optimal \mbox{worst-case} and the optimal average packet rates are achieved by activating overlapping paths. Moreover, operating overlapping paths can provide additional benefits, e.g., decreasing the variance of the achieved packet rate.
\item We present a heuristic path selection algorithm that modifies Dijkstra's shortest path algorithm and combines it with the LP that characterizes the optimal average packet rate so as to proactively and efficiently select (in polynomial time in $N$) resilient paths with high packet rates in a mmWave network with arbitrary topology. We pair this algorithm with a DRL-based online approach for decentralized adaptation to link failures, and we develop a \textit{hybrid} scheduling algorithm that ensures a target packet rate.
To the best of our knowledge, this is the first time that a DRL-based hybrid algorithm is used for online optimization of packet rates of multiple paths in mmWave networks. 



\item The performance of the proposed algorithm is evaluated against alternative algorithms in volatile environments. The proposed scheduling algorithm robustly adapts to link failures over different topologies, channel and blockage realizations. As our evaluation results indicate, it offers a superior performance to alternative algorithms.
\end{enumerate}

\subsection{Related Work} 
A multitude of works in the literature propose relay selection schemes~\cite{dimas2019cooperative,Yan18,7565000,Ibrahim21}, but focus on selecting the single path that has the highest Signal-to-Noise Ratio (SNR), and thus do not offer resilience to blockage. Several works study scheduling and routing in wireless networks by exploring multi-path diversity~\cite{Zhou12,Rocher2009,Shillingford2008,YeungBook}. 
However, these works focus on 
traditional
wireless networks and they do not consider the scheduling constraints of mmWave communications. Some studies on mmWave communications have focused on profiling the distribution of the Signal-to-Interference-plus-Noise Ratio
(SINR) in random environments both in cellular and ad-hoc network settings~\cite{Bai2014,Thornburg2015}. However, these works consider communication over a single-hop either between ad-hoc nodes or in a cellular system between a base station and a user equipment, and they do not characterize the optimal \mbox{worst-case} or the optimal average packet rates as we do. 
There exist studies that address path selection and rate allocation problems in multi-hop mmWave networks~\cite{Garcia15,Hu17,Sahoo17,Yan18,Vu19}. However, these works either do not consider reliability or they rely on the knowledge of channel state information (CSI) (or network topology) which can be prohibitively costly to acquire. In this paper, we develop a resilient scheduling algorithm that does not require knowledge of network topology or link capacities frequently, thus it is well suited to volatile environments.

There are studies that aim to reduce link outages in mmWave networks by taking reactive approaches~\cite{Jeong,Barati,Abari}. However, such a reactive mechanism adds the complexity of identification and adaptation, as well as feedback latency.
In~\cite{MineMilcom}, to achieve resilience to link and node failures, the authors explored a DRL algorithm, which adapts the information flow through the mmWave network, without using knowledge of the link capacities or network topology. However, this proposed approach also reactively adapts to network disruptions, and the authors do not propose a path selection algorithm for the RL approach. Thus, the algorithm is sensitive to initially selected paths and channel conditions. In this paper, we develop and leverage proactive transmission mechanisms to select proactively resilient paths with high packet rates for our DRL-based hybrid scheduling algorithm to reduce the complexity of identification and adaptation, feedback latency and sensitivity to network topology and channel conditions. Our algorithm robustly adapts to link failures in different topologies and channel conditions. 

Several works proposed proactive approaches that constantly track users using side-channel information~\cite{SurMobicom,Nitsche} or external sensors~\cite{Haider,Wei,Va}. These solutions have limited accuracy, and possibly require sensitive information, such as user location~\cite{Wei,Va}. In~\cite{MineISIT22}, authors leverage scheduling properties of mmWave links, as well as the blockage asymmetry, to design schemes that achieve the average and the \mbox{worst-case} approximate capacities and proactively offer high resiliency. In this paper, we build on the scheme that characterizes the optimal average packet rate to develop a heuristic path selection algorithm, and design a hybrid scheduling algorithm that gracefully adapts to network disruptions.

\noindent \textbf{Paper Organization. } Section~\ref{section_model} provides background on the 1-2-1 network model for mmWave networks and on DRL methods. Section~\ref{section_proactive} presents proactive transmission mechanisms for centralized adaptation to link failures. Section~\ref{section_hybrid} presents a hybrid scheduling mechanism that combines the proactive transmission method with a DRL based reactive mechanism for decentralized adaptation to link failures. Section~\ref{section_evaluation} presents the evaluation results of the algorithm. 

\section{System Model and Background}\label{section_model}
\textit{Notation.} With $[n_1:n_2]$, we denote the set of integers from $n_1$ to $n_2 \geq n_1$. $\varnothing$ is the empty set and $|\cdot|$ denotes the cardinality for sets. $\mathbb{E} [\cdot]$ and $\mathbb{V} (\cdot)$ denote the expectation and the variance of a random variable, respectively. $H (\cdot)$ denotes the entropy of a random variable. 
\subsection{Gaussian 1-2-1 Networks}
We consider the Full-Duplex (FD) Gaussian 1-2-1 network model that was introduced in~\cite{ezzeldin} to study the information-theoretic capacity of multi-hop mmWave networks. In an $N$-relay Gaussian FD 1-2-1 network model, $N$ relays assist the communication between a source node (node $0$) and a destination node (node $N+1$). Each node in the network can simultaneously transmit and receive by using a single transmit beam and a single receive beam. At any particular instance, a node can transmit to at most one node and it can receive from at most one node. In order for two nodes to communicate, they need to steer their beams towards each other so as to
activate a link (edge) that connects them. 

\noindent
\textbf{Capacity of Gaussian FD 1-2-1 networks.} In~\cite{ezzeldin}, it was shown that the unicast capacity of an $N$-relay Gaussian FD 1-2-1 network can be approximated to within an additive gap that only depends on the number of nodes in the network. In particular, the following LP
was proposed to compute the unicast approximate capacity and its optimal schedule in polynomial-time,
\begin{align}\label{capacity_paths}
\begin{array}{llll}
&\ {\rm{P1:}}\  \displaystyle \widebar{\mathsf{C}} = \underset{x_p, p \in \mathcal{P}}{\max} \displaystyle\sum_{p \in \mathcal{P}} x_p \mathsf{C}_p   & &\\
& ({\rm P1}a) \ x_p \geq 0, & \forall p  \in  \mathcal{P}, &\\
& ({\rm P1}b) \ \displaystyle\sum_{p \in \mathcal{P}_i}  x_p f^p_{p\pnext(i),i} \leq 1, & \forall i \in  [0 : N], & \\
& ({\rm P1}c) \ \displaystyle\sum_{p \in \mathcal{P}_i} x_p f^p_{i,p\pprev(i)} \leq 1, & \forall i  \in  [1:N+1],  &
\end{array}
\end{align}
where: (i) $\widebar{\mathsf{C}}$ is the approximate capacity; 
(ii) $\mathcal{P}$ is the collection of all paths connecting the source to the destination; (iii) $\mathsf{C}_p$ is the capacity of path $p$; (iv) $\mathcal{P}_i \subseteq \mathcal{P}$ is the set of paths that pass through node $i$ where $i \in [0:N+1]$; (v) $p\pnext(i)$ (respectively, $p\pprev(i)$) is the node that follows (respectively, precedes) node $i$ in path $p$; (vi) $x_p$ is the fraction of time path $p$ is used; and (vii) $f^p_{j,i}$ is the optimal activation time for the link of capacity $\ell_{j,i}$ when path $p$ is operated, i.e., $   f^p_{j,i} = {\mathsf{C}}_p/\ell_{j,i}.$
Here, $\ell_{j,i}$ denotes the capacity of the link going from node $i$ to node $j$ where $(i,j) \in  [0:N]\times[1:N+1]$.

Although the number of variables in LP $\rm{P1}$ (particularly, the number of paths) can be exponential in the number of nodes, this LP can be solved in polynomial-time through an equivalent LP
as proved in~\cite{ezzeldin}. We refer the interested reader to~\cite{ezzeldin} for a more detailed description.

\begin{remark}
In LP $\rm{P1}$, the beam scheduling enables 
the sharing of the traffic across multiple paths, 
both over space and time without considering resilience to link and node failures. Our aim is to determine which paths to use and how to schedule them with the two-fold objective of ensuring resilience against link blockages and achieving a high end-to-end packet delivery rate.
\end{remark}
In mmWave networks, the probability of network disruptions can be highly asymmetric, yet known in advance through accurate models~\cite{Jain,Samuylov16,Gapeyenko17,Wang17,Raghavan19}. In particular, expressions for the blockage rate of Line-of-Sight (LOS) links were derived by modeling the arrival process of blockers as a Poisson Point Process (PPP). The blockage rate $\alpha_{j,i}$ of the link from node $i$ to node $j$ is
\begin{equation}\label{block_prob}
    \alpha_{j,i} = \lambda_{j,i}d_{j,i},
\end{equation}
where: (i) $d_{j,i}$ is the distance between node $i$ and node $j$; and (ii) $\lambda_{j,i}$ is proportional to the blocker density and velocity, as well as to the heights of the blocker, receiver and transmitter.
In this paper, we are interested in characterizing the optimal \mbox{worst-case} and optimal average packet rates, as defined below.
\begin{definition}
The optimal packet rate is the fraction of received packets under an optimal beam schedule. 
\end{definition}
\begin{remark}
When there are no packet losses, the optimal packet rate is equal to the approximate capacity in~\eqref{capacity_paths} . However, in this paper we consider link blockages and hence, we talk about packet rates.
\end{remark}
We consider a permanent blockage (failure) model, where the link of capacity $\ell_{j,i}$ is blocked with probability $p_{j,i}$ and it is not blocked with probability $(1-p_{j,i})$. This model is different from the erasure channel model where a packet is blocked with probability $p_{j,i}$ at every channel use. In the following example, we show that the optimal solution for the erasure channel model does not necessarily give the optimal solution for the permanent blockage model that we consider.

\noindent\textit{Example 1}. Consider the network in Fig.~\ref{permanent_block} in which link capacities are $\ell_{2,0} = 4,~\ell_{3,2} = 12,~\ell_{1,0} = \ell_{3,1} = 3,~ \ell_{4,3} = 6$ and the link blockage probabilities are zero except for $p_{3,2} = 2/3$. There are two paths connecting the source (node $0$) to the destination (node $4$), particularly: $p_1: 0 \rightarrow 1 \rightarrow 3\rightarrow 4$ and $p_2: 0 \rightarrow 2 \rightarrow 3\rightarrow 4$.
\begin{figure}
	\centering
    \includegraphics[width=0.23\columnwidth]{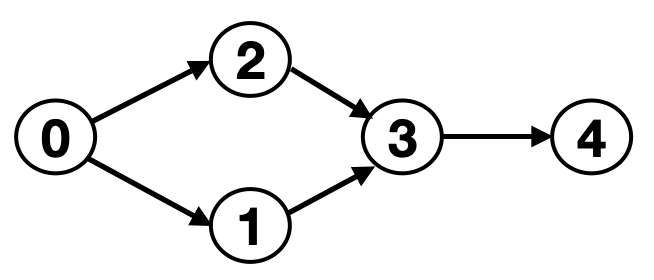}
     \caption{An mmWave network example with $N=3$ relay nodes.}
     \label{permanent_block}
     \vspace{-0.15in}
\end{figure}
In Fig.~\ref{permanent_block}, for the erasure channel model, we can simply replace the link capacities $\ell_{j,i}$ with the average link capacities $(1-p_{j,i})\ell_{j,i}$ and solve LP $\rm{P1}$ in~\eqref{capacity_paths} to find the optimal average packet rate. The optimal solution of LP $\rm{P1}$, i.e., the optimal schedule would activate path $p_2$ because $p_2$ has a higher packet rate than $p_1$. However, for the permanent blockage model that we consider, two scenarios can happen, namely:
(1) the link with capacity $\ell_{3,2}$ is blocked and hence, path $p_2$ is blocked with probability $2/3$; and (2) none of the links is blocked with probability $1/3$. The optimal schedule would activate path $p_1$ because even though the packet rates of $p_1$ and $p_2$ are similar, there is a high probability that $p_2$ can get blocked.
\hfill $\square$

\subsection{Deep Reinforcement Learning}
We here provide some background on DRL and on the state-of-the-art SAC algorithm, which we will use in the design of our communication protocols.
In RL, an agent observes an environment and interacts with it. At time step $t$, the agent at state $\mathbf{s}_t$ takes the action $\mathbf{a}_t$ and it moves to the next state $\mathbf{s}_{t+1}$, while receiving the reward $r(\mathbf{s}_t,\mathbf{a}_t)$. In RL settings, states represent the environment and the set of all states is called the state space $\mathcal{S}$. Actions are chosen from an action space $\mathcal{A}$, where we use $\mathcal{A}\left(\mathbf{s}_t\right)$ to denote the set of possible (valid) actions at state $\mathbf{s}_t$. Rewards are numerical values given to the agent according to its actions and the aim of the agent is to maximize the long-term cumulative reward~\cite{10.5555/3312046}. At each time step, the agent follows a policy $\pi\left(\cdot \;\middle\vert\; \mathbf{s}_t\right)$ and chooses an action from the action space. Policy $\pi$ is a distribution (potentially deterministic) over actions given the current state $\mathbf{s}_t$. One of the ways to determine how useful a particular action is at a given state is to evaluate the Q-function, which computes the expected cumulative reward for an action and a state under policy $\pi$. We consider episodic RL, where the agent interacts with the environment for a finite horizon $T$ to maximize its long term cumulative reward.
In a lot of RL applications, from games to robotic control, we consider continuous state and action spaces which render classical tabular methods prohibitively inefficent. Thus, function approximators and model-free RL techniques are  employed to deal with the shortcomings~\cite{Arulkumaran_2017}. Although such techniques can be successful on challenging tasks, they suffer from two major drawbacks: high sample complexity and sensitivity to hyperparameters. Off-policy learning algorithms are proposed to improve sample efficiency, but they tend to experience stability and convergence issues particularly in continuous state and action spaces. 

The state-of-the-art SAC algorithm, an off policy deep RL algorithm,  was proposed in~\cite{haarnoja2018soft} to improve the exploration and stability. Since large state-action spaces require function approximators, SAC  uses approximators for the policy and Q-function. In particular, the algorithm uses five parameterized functional approximators: policy function $(\phi)$; soft Q-functions $(\theta_1$ and $\theta_2)$; and target soft Q-functions $(\bar{\theta}_1$ and $\bar{\theta}_2)$. The aim is to maximize the following objective function
\begin{align}
J(\pi) = \sum_{t = 0}^{T}\mathbb{E}\left[r\left(\mathbf{s}_t,\mathbf{a}_t\right)+\gamma H\left(\pi\left(\cdot \;\middle\vert\; \mathbf{s}_t\right)\right)\right],
\label{sac_objective}
\end{align}
where: (i) $T$ is the horizon; and (ii) the temperature parameter $\gamma$ indicates the relative importance of the entropy term to the reward. The entropy term  enables the SAC algorithm to achieve improved exploration, stability and robustness~\cite{haarnoja2018soft}. 

\section{Proactive Transmission Mechanisms for Centralized Adaptation}\label{section_proactive}
In this section, we aim to build scheduling mechanisms that provide proactive resilience against link blockages/failures in a mmWave network with arbitrary topology. In particular, we are interested in characterizing both the optimal {\em \mbox{worst-case}} and the optimal {\em average} packet rates.

In the {\em worst case}, we assume that $k_{\ell}\in [0:|\mathcal{E}|]$  links fail, where $\mathcal{E}$ is the set of network links. We highlight that 
the set of the $k_{\ell}$ blocked links is not known, i.e., only the link blockage probabilities are known. Under such assumptions, we can find the optimal {\em \mbox{worst-case}} packet rate by solving 
$\rm{P1}$ in~\eqref{capacity_paths} with the objective function modified as follows (the constraints are the same as those in LP $\rm{P1}$):
\begin{itemize}
    \item {\em Optimal Worst-Case Packet Rate}: 
    \begin{align}
    \label{eq:WSCap}
    \underset{x_p, p \in \mathcal{P}}{\max} \ \underset{a \in \mathcal{B}}{\min} \ \underset{p \in \mathcal{P}^{(a)}}{\sum} x_p \mathsf{C}_p,
    \end{align}
\end{itemize}
where: (i) $\mathcal{B}$ is the set of all combinations of $k_\ell$ links from the $|\mathcal{E}|$ links; and (ii) $\mathcal{P}^{(a)}$ is the set of unblocked paths when the $k_\ell$ links in $a \in \mathcal{B}$ are blocked. 

In the {\em average case}, we remove the assumption of $k_\ell$-size failure patterns, and we find the optimal average packet rate over {\em all} failure patterns by solving the LP $\rm{P1}$ in~\eqref{capacity_paths}, where the objective function is now modified as follows (the constraints are the same as those in LP $\rm{P1}$):
\begin{itemize}
\item {\em Optimal Average Packet Rate:}
    \begin{align}
    \label{eq:AveCap}
    \underset{x_p, p \in \mathcal{P}}{\max} \ \underset{p \in \mathcal{P}}{\sum} x_p \mathsf{C}_p \left( \displaystyle \prod_{(j,i) \in \mathcal{E}_p} \left( 1-p_{j,i} \right ) \right ),
    \end{align}
\end{itemize}
where $\mathcal{E}_p$ is the set of links in path $p \in \mathcal{P}$.

We highlight that solving $\rm{P1}$ in~\eqref{capacity_paths} with one of the above two objective functions finds a schedule that achieves either the optimal \mbox{worst-case} or the optimal average packet rate.
\begin{remark}
LP $\rm{P1}$ in~\eqref{capacity_paths} with the objective function in~\eqref{eq:WSCap} can be equivalently formulated with a number of variables polynomial in $N$ by taking the same steps as in~\cite{ezzeldin}. 
Thus, if $|\mathcal{B}|$ in~\eqref{eq:WSCap} is a polynomial function of $N$, the optimal \mbox{worst-case} packet rate and an optimal schedule for it can be computed in polynomial-time in $N$. For the optimal average packet rate, we have an exponential number of variables (the number of paths) and finding whether an equivalent formulation with a polynomial number of variables in $N$ exists, is an interesting open direction, which is currently under investigation. As pointed out in Example~1, simply replacing each link capacity with the average link capacity and solving LP $\rm{P1}$ does not find the optimal average packet rate.
\end{remark}
As we show through the following simple example, LP $\rm{P1}$ in~\eqref{capacity_paths} without the modifications in~\eqref{eq:WSCap} and~\eqref{eq:AveCap} does not ensure resilience against link failures/blockages.

\noindent \textit{Example 2.} Consider the network with in Fig.~\ref{basevsworst} when $k_{\ell} = 1$ link is blocked.
\begin{figure*}
     \centering
     \begin{subfigure}[b]{0.25\textwidth}
         \centering
         \includegraphics[width=\textwidth]{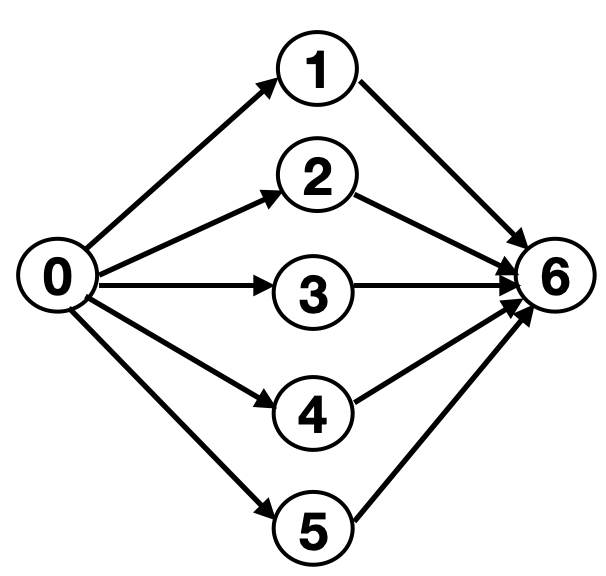}
         \caption{}
         \label{basevsworst}
     \end{subfigure}
     \begin{subfigure}[b]{0.25\textwidth}
         \centering
         \includegraphics[width=\textwidth]{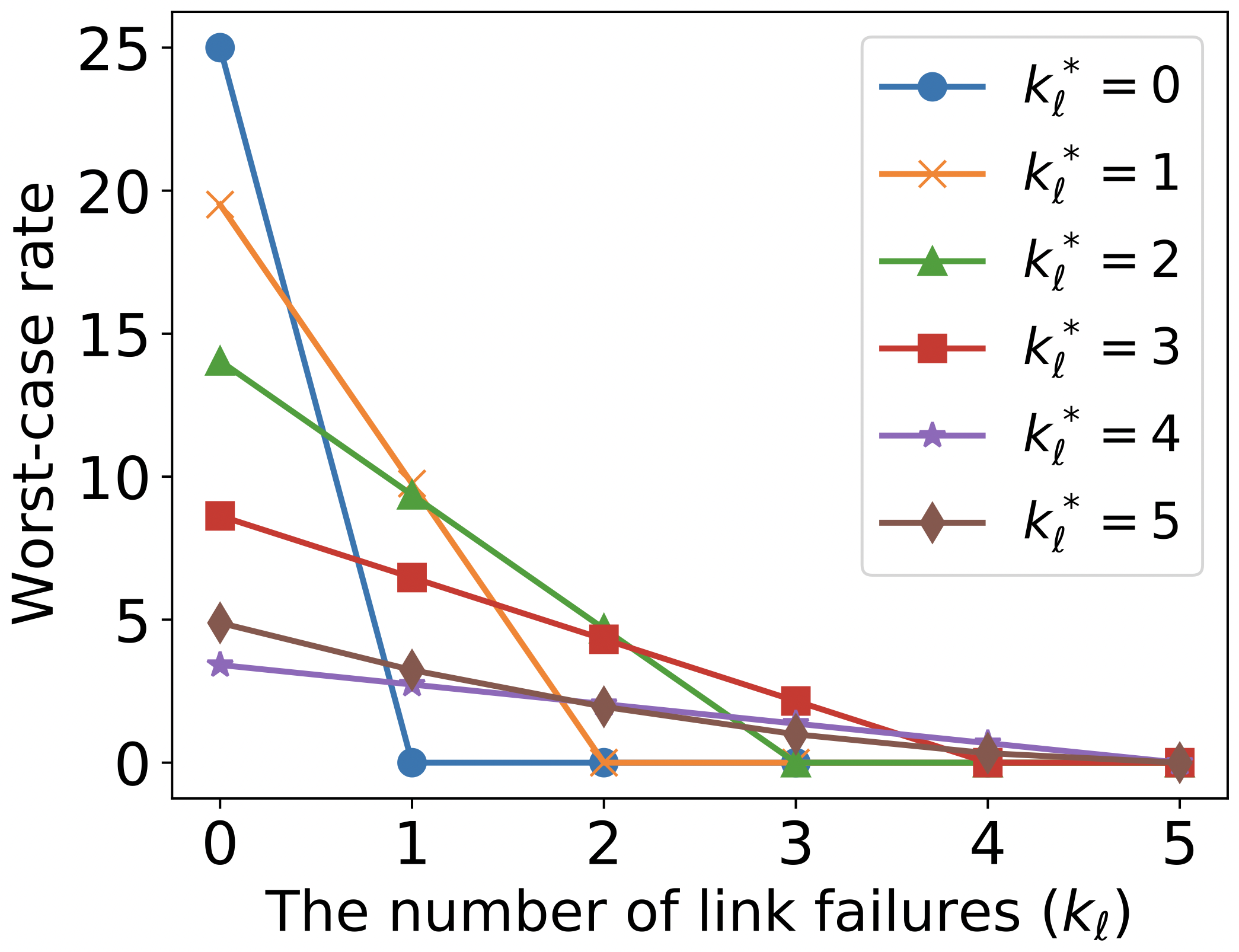}
         \caption{}
         \label{tradeoff_worst}
     \end{subfigure}
        \caption{An mmWave network example with resilience-optimal packet rate trade-off curve.}
        \vspace{-0.1in}
\end{figure*}
There exist five paths connecting the source (node $0$) to the destination (node $6$), particularly $p_1: 0 \rightarrow 1 \rightarrow 6,~p_2: 0 \rightarrow 2 \rightarrow 6,~p_3: 0 \rightarrow 3 \rightarrow 6,~ p_4: 0 \rightarrow 4 \rightarrow 6$, and $p_5: 0 \rightarrow 5 \rightarrow 6$  where the path capacities are $c^2,4c^2,9c^2,16c^2$ and $25c^2$, respectively. Here, $c\geq1$ is a constant and the capacities of the links which are on the same path are equal. The optimal solution of LP $\rm{P1}$ activates only path $p_5$ to achieve the approximate capacity $25c^2$. This solution does not ensure resilience as the \mbox{worst-case} packet rate for $k_\ell = 1$ is $0$ (i.e., when any link on $p_5$ is blocked). However, for the worst case, resilience is ensured by the schedule that maximizes the objective function in~\eqref{eq:WSCap}, which activates the strongest two paths such that an equal packet rate is sent through them. Particularly, it activates $p_4$ and $p_5$ for $25/41$ and $16/41$ fractions of time (inversely proportional to the path packet rate), respectively. The optimal \mbox{worst-case} packet rate is equal to $9.76c^2$. As we increase the value of $c$, the gain we obtain by solving for the objective function in~\eqref{eq:WSCap} increases. A similar result is obtained for the optimal average packet rate. We assume that the path capacities are $c,2c,3c,4c$ and $5c$ where $c\geq1$, and the capacities of the links on the same path are equal. The link blockage probabilities are all equal to $1/10$ except for the links in $p_5$ for which the
blockage probabilities are equal to $2/3$. The average packet rate achieved by LP $\rm{P1}$ is $0.56c$ and the optimal average packet rate in~\eqref{eq:AveCap} is equal to $3.24c$. As we increase the value of $c$, the gain we obtain by solving for the objective function in~\eqref{eq:AveCap} increases. \hfill $\square$

As \textit{Example~2} illustrates, the optimal \mbox{worst-case} packet rate over a given topology can significantly depend on the level of resilience (captured by the value of $k_\ell$) that we want to offer. In Fig.~\ref{tradeoff_worst}, we present the resilience-optimal packet rate \mbox{trade-off} curve for the network in Fig.~\ref{basevsworst} for $c = 1$.
The resilience-optimal packet rate \mbox{trade-off} for a mmWave network is the information theoretic Pareto-optimal vector of packet rates that can be achieved despite any $k_\ell \in [0:|\mathcal{E}|]$ link failures. In Fig.~\ref{tradeoff_worst}, for each $k_\ell^* \in [0:5]$, we find a schedule that achieves the optimal \mbox{worst-case} packet rate in~\eqref{eq:WSCap} when we assume that {\em any} $k_\ell^*$ network links fail (the optimal \mbox{worst-case} packet rate for $k_\ell^* \geq 6$ link failures is equal to $0$). Under this optimal schedule, we then find the achieved \mbox{worst-case} packet rate as the number of link failures increases from $0$ to $5$. For example, the optimal schedule for $k_\ell^* = 2$ activates the three strongest paths $p_3,p_4$ and $p_5$; hence, for $k_\ell^* = 2$, as $k_\ell$ increases in Fig.~\ref{tradeoff_worst}, the achieved \mbox{worst-case} packet rate decreases and becomes $0$ for $k_\ell \geq 3$. Fig.~\ref{tradeoff_worst} shows that there is a \mbox{trade-off} between \mbox{packet-rate} and resilience, i.e., guaranteeing a certain amount of (worst case) resilience to say $k_\ell^* = 1$ link blockage, may come at the cost of a lower packet rate when no link is blocked (i.e., $k_\ell = 0$).

\begin{remark}
In Fig.~\ref{tradeoff_worst}, the optimal schedule for $k_\ell^*$ link failures activates $k_\ell^*+1$ paths. However, different strategies might be required for different values of $k_\ell^*$ depending on the number of paths and path capacities. For example, if we add $5$ more edge-disjoint paths in the network in Fig.~\ref{basevsworst} with capacities $c^2i^2$ for $i \in [6:10]$, the optimal schedule for $k_\ell^* = 1$ link failure activates the strongest $3$ paths, and
the optimal schedule for $k_\ell^* = 2$ activates the strongest $5$ paths. Moreover, more paths are activated if the path capacities are closer to each other: for equal path capacities, the optimal solution always (i.e., independently of $k_\ell^\star$) activates all the paths. In particular, the optimal solution finds a schedule that allocates
an equal packet rate through each path. This leads to a trade-off: if a higher number of paths are activated, every link failure in the worst case decreases the achieved packet rate less. 
However, activating a higher number of paths results in operating lower capacity paths for a longer time, and higher capacity paths for a shorter time. The optimal strategy is determined based on this trade-off. 
\end{remark}
We now conclude this section with a few observations regarding the optimal \mbox{worst-case} and optimal average packet rates in~\eqref{eq:WSCap} and~\eqref{eq:AveCap}, respectively. We start with
the  Lemma~\ref{lem:LinPaths}, which shows that the {\em optimal average packet rate}  can always be achieved by using only a linear number (in $N$) of paths. We note that the following lemma is a generalization of~\cite[Lemma 9]{ezzeldin}.
\begin{lemma}
\label{lem:LinPaths}
For any $N$-relay Gaussian FD 1-2-1 network, the optimal average packet rate can always be achieved by activating at most $2N+2$ paths in the network.
\end{lemma}
\begin{proof}
The LP with objective function in~\eqref{eq:AveCap} and constraints in LP $\rm{P1}$ in~\eqref{capacity_paths}
is bounded and thus, there always exists an optimal vertex. In particular, each vertex of this LP satisfies at least $|\mathcal{P}|$ inequality constraints with equality among $(\rm{P1}a),~(\rm{P1}b)$ and $(\rm{P1}c)$. $(\rm{P1}b)$ and $(\rm{P1}c)$ combined represent $2N+2$ constraints, thus we have at least $|\mathcal{P}|-2N-2$ constraints in $(\rm{P1}a)$ that are satisfied with equality. Therefore, at least $|\mathcal{P}|-2N-2$ paths do not operate. 
\end{proof}
\begin{remark}
A similar result as in Lemma~\ref{lem:LinPaths} does not hold for the optimal \mbox{worst-case} packet rate. To see this, consider the case when $k_\ell = 2N+2$, which might cause $2N+2$ paths to be blocked, hence leading to a zero packet rate
in the worst case. Indeed, expressing  (\ref{eq:WSCap}) as an LP introduces additional constraints resulting in a different  vertex structure than the problem in~\eqref{eq:AveCap}.
\end{remark}
Lemma~\ref{lem:LinPaths} shows that at most $2N+2$ paths suffice to achieve the optimal average packet rate. As we show through the next example, a much smaller number of paths might indeed be sufficient to characterize the optimal average packet rate.

\noindent \textit{Example~3.} Consider the network in Fig.~\ref{linear_paths} with unitary link capacities, and the link blockage probabilities are all equal to $1/5$.
\begin{figure}
	\centering
     \includegraphics[width=0.35\columnwidth]{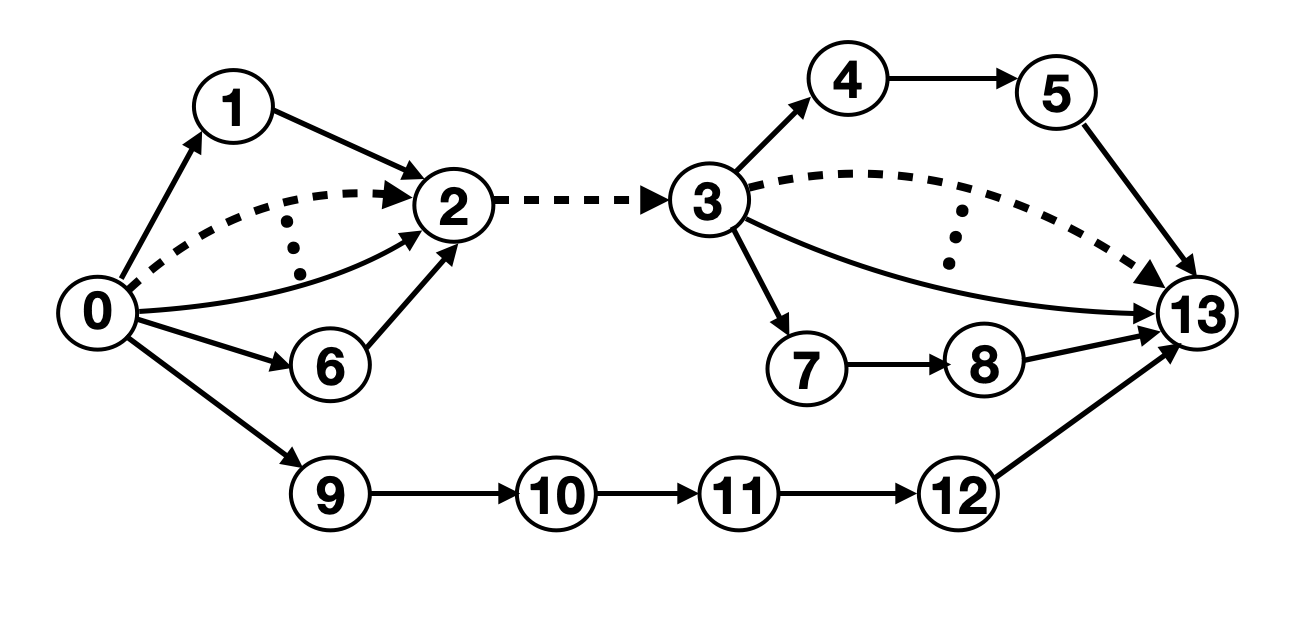}
       \vspace{-0.2in}
     \caption{An mmWave network example with $N=12$ relay nodes for Lemma~\ref{lem:LinPaths}.}
     \label{linear_paths}
     \vspace{-0.15in}
\end{figure}
There exist $2705$ paths connecting 
node $0$ to node $13$. The dots from node $0$ to node $2$ (and similarly, the dots from node $3$ to node $13$) in Fig.~\ref{linear_paths} represent $50$ links connecting them. From Lemma~\ref{lem:LinPaths} we know that operating $2N+2 = 26$ paths suffices to achieve the optimal average packet rate.
However, there is no need to activate all of these paths to achieve the optimal average packet rate of $64/125$. Instead, it suffices to activate only $1$ of these paths, for instance
$0 \rightarrow 2 \rightarrow 3 \rightarrow 13$ (highlighted with dashed lines in Fig.~\ref{linear_paths}). \hfill $\square$

A question that naturally arises in finding a schedule that provides resilience to link failures, and achieves the optimal \mbox{worst-case} or optimal average packet rate is the following: What are the {\em best} paths to use? Or in other words, are there any intrinsic properties of the paths that should be leveraged? The next theorem provides an answer to these questions.

\begin{theorem}\label{results}
For an $N$-relay Gaussian FD 1-2-1 network with arbitrary topology,

\noindent {\rm{(P1)}} When the link capacities are all \textit{equal}, the optimal \mbox{worst-case} and the optimal average packet rates can always be achieved by activating only edge-disjoint paths.

\noindent {\rm{(P2)}} When the link capacities are \textit{unequal}, there exist network topologies for which the optimal \mbox{worst-case} and the optimal average packet rates are achieved by activating overlapping paths.
\end{theorem}

\begin{proof}
The proof of~\rm{(P1)} is delegated to the Appendix. We here focus on proving \rm{(P2)}. Towards this end, we consider the network in Fig.~\ref{overlapping}.
There exist $5$ paths connecting the source (node $0$) to the destination (node $13$), particularly: $p_1: 0 \rightarrow 1 \rightarrow 2\rightarrow 3 \rightarrow 4 \rightarrow 5 \rightarrow 13,~p_2: 0 \rightarrow 1 \rightarrow 2\rightarrow 3 \rightarrow 7 \rightarrow 8 \rightarrow 13,~ p_3: 0 \rightarrow 6 \rightarrow 2\rightarrow 3 \rightarrow 4 \rightarrow 5 \rightarrow 13,~p_4: 0 \rightarrow 6 \rightarrow 2 \rightarrow 3 \rightarrow 7 \rightarrow 8 \rightarrow 13,~p_5: 0 \rightarrow 9 \rightarrow 10 \rightarrow 11 \rightarrow 12 \rightarrow 13$. The link capacities are assumed to be $\ell_{1,0} = \ell_{2,1} = \ell_{4,3} = \ell_{5,4} =\ell_{13,5} = \ell_{8,7}= 1,~
 \ell_{9,0} = \ell_{10,9} = \ell_{11,10} = \ell_{12,11} = \ell_{13,12} = 4,~
\ell_{6,0} = \ell_{2,6}  = \ell_{3,2}  = \ell_{7,3}  = \ell_{13,8} = 2.$
We first consider the {\em \mbox{worst-case}} scenario for $k_{\ell} = 1$ (i.e., only one link is blocked). In an optimal solution of
the LP with objective function in~\eqref{eq:WSCap} and constraints in the LP $\rm{P1}$ in~\eqref{capacity_paths},
three paths $p_3,~p_4$ and $p_5$ are activated with activation times $0.2,1$ and $0.3$, respectively. 
The optimal \mbox{worst-case} packet rate is $1.2$. As seen in Fig.~\ref{overlapping}, $p_3$ and $p_4$ share the edges with capacities $\ell_{6,0},~\ell_{2,6}$ and $\ell_{3,2}$. We note that none of the feasible solutions that consist of only edge-disjoint paths reaches the same or a higher \mbox{worst-case} packet rate.
We now consider the optimal average packet rate  over the same network in Fig.~\ref{overlapping} with the same link capacities except for the link capacities in $p_5$ that are now assumed to be unitary. The link blockage probabilities are assumed to be equal to $1/5$ except for the links in $p_5$ for which the blockage probabilities are assumed to be equal to $1/3$. In this case, we observe a similar situation as in the \mbox{worst-case} scenario. 
In particular, an optimal solution of the LP with objective function in~\eqref{eq:AveCap} and constraints in the LP $\rm{P1}$ activates three overlapping paths, namely $p_2,p_3$ and $p_4$ with equal activation times of $0.5$, and optimal average packet rate  is $0.39$. We note that none of the feasible solutions consisting of only edge-disjoint paths reaches the same or a higher average packet rate. 
This concludes the proof of Theorem~\ref{results}.
\end{proof}
\begin{figure}
	\centering
     \includegraphics[width=0.35\columnwidth]{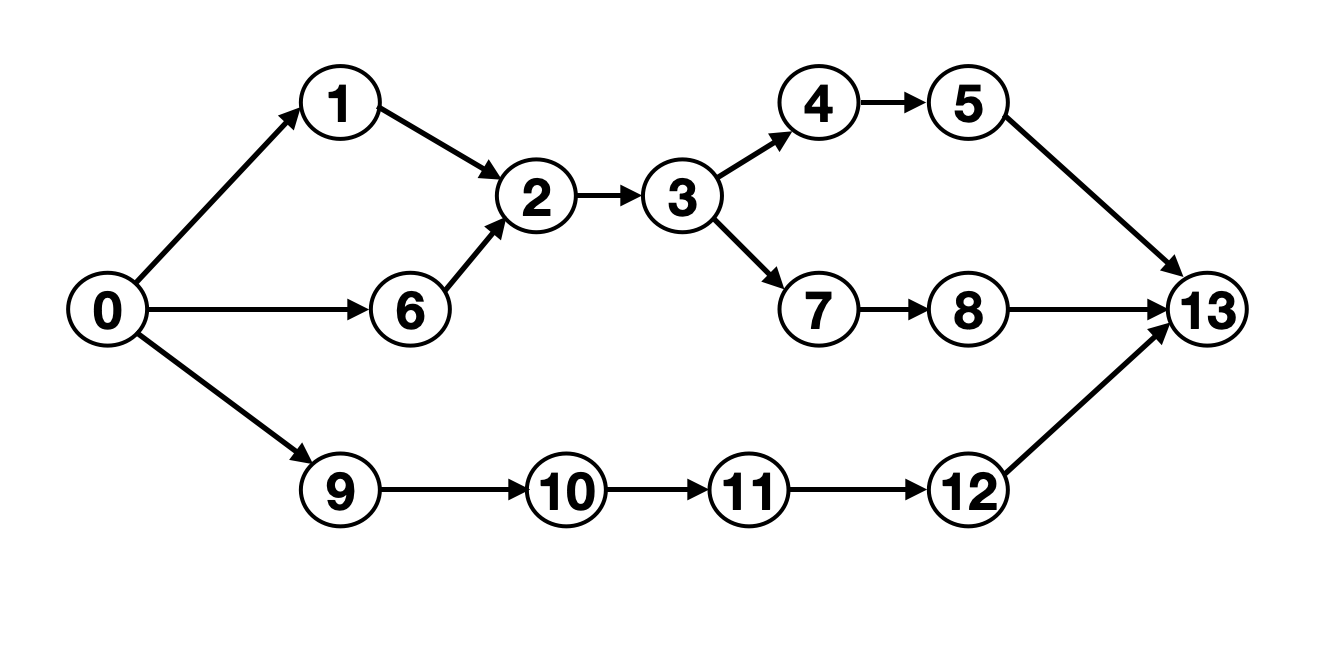}
     \vspace{-0.25in}
     \caption{Example network with overlapping paths.}
     \label{overlapping}
     \vspace{-0.15in}
\end{figure}
Intuitively, activating edge-disjoint paths would provide higher resilience against link failures since a higher number of paths can get blocked as the number of shared edges increases. However, Theorem~\ref{results} shows that activating overlapping paths may provide higher resilience against link failures when the link capacities are unequal. One intuitive explanation for this is that, even though overlapping paths share links, if the shared links have low blockage probability and high link capacity, operating the overlapping paths might still result in higher packet rates.

We conclude this section with two remarks that show interesting consequences of Theorem~\ref{results}.
\begin{remark}
The gain between the optimal \mbox{worst-case} packet rate (obtained by using overlapping paths) and the \mbox{worst-case} packet rate obtained by using only edge-disjoint paths can be arbitrarily large. As an example, consider the network in Fig.~\ref{overlapping} with $k_{\ell}=1$ and increase the link capacities in $p_4$ (except for $\ell_{8,7}$) and $p_5$. The optimal \mbox{worst-case} packet rate (obtained by using overlapping paths) approaches $2$. The \mbox{worst-case} packet rate obtained by only using edge-disjoint paths approaches $1$, thus the gain approaches~$1$, which is the largest gain that we can have in this network.
In a similar manner, the gain between the optimal average packet rate
 (obtained by using overlapping paths) and the average packet rate obtained by using only edge-disjoint paths can be arbitrarily large. In Fig.~\ref{overlapping}, as we increase the link capacities in $p_4$ (except for $\ell_{8,7}$) as well as the blockage probabilities of the links in $p_5$, and decrease the blockage probabilities of the remaining links, the optimal average packet rate (obtained by using overlapping paths) approaches $2$. The average packet rate obtained by only using edge-disjoint paths approaches $1$, thus the gain approaches $1$, which is the largest gain that we can have in this network. 
\end{remark}

\begin{remark}
Theorem~\ref{results} shows that if the solution of the LP with objective function in~\eqref{eq:AveCap} and constraints as in the LP $\rm{P1}$ activates overlapping paths for a network with equal link capacities, then there exists another solution consisting of only edge-disjoint paths. However, activating overlapping paths can still provide additional benefits, such as decreasing the variance of the achieved rate as illustrated next.
\end{remark}
\noindent \textit{Example~4}. We consider the network in Fig.~\ref{overlapping} without $p_5$. The link capacities are assumed to be equal to $2$ and the blockage probabilities are all equal to $1/5$. We consider the optimal average packet rate given by solving the LP with objective function in~\eqref{eq:AveCap} and constraints as in LP $\rm{P1}$.
Optimal solutions that only operate edge-disjoint paths activate only one of the paths with an activation time of $1$. The optimal average packet rate is $\mathbb{E}[\mathsf{R}] = 0.52$ and the variance $\mathbb{V}(\mathsf{R}) = 0.77$ where $\mathsf{R}$ denotes the achieved rate. Another optimal solution activates all the four paths $p_1,p_2,p_3,p_4$ all with activation time equal to $0.25$. In this case, $\mathbb{V}(\mathsf{R}) = 0.38$, which is much lower than $0.77$. \hfill $\square$


\section{Hybrid Scheduling Algorithm for Decentralized Adaptation}\label{section_hybrid}
The proactive mechanisms proposed in Section~\ref{section_proactive} rely on a centralized knowledge of the network link capacities to find an optimal schedule. In this section, we develop a hybrid scheduling algorithm that combines our proactive mechanisms with an online approach for decentralized adaptation to link failures. Our algorithm relies on the interaction between a DRL agent and the network. In particular,
we assume that the channel coefficients (and, as a consequence, the link capacities) are unknown and they can change over time. Our aim is to attain a certain desired packet rate $R^\star$ by using a small subset of the possible paths $\mathcal{P}_k$, where $\mathcal{P}_k \subseteq \mathcal{P}$ and $\left|\mathcal{P}_k\right| = k$. Thus, the input and output size of the RL algorithm is equal to $k$. The number $k$ affects the complexity of our algorithm. An implicit requirement is that, the selected set of paths $\mathcal{P}_k $ can jointly support the desired rate. 
The reason we aim to reach a  desired packet rate, instead of  the full network optimal packet rate, is that achieving the optimal packet rate may require the use of a  large number of paths, which can increase the dimension of the problem substantially and render the RL problem infeasible for large networks. We next propose a path selection algorithm for our RL approach to create the set $\mathcal{P}_k$.

\subsection{Path Selection Algorithm}
\label{sec:PathSelection}
As highlighted in Section~\ref{section_proactive}, solving the LP with objective function as in~\eqref{eq:AveCap} and constraints as in the LP $\rm{P1}$ in~\eqref{capacity_paths} finds a schedule that provides resilience against link failures while achieving a high packet rate. We here leverage this proactive transmission mechanism to select a subset of paths for the RL algorithm, i.e., in order to create the set $\mathcal{P}_k$. 

We start by noting that, over a mmWave network with arbitrary topology, there might exist an exponential number of paths (in $N$) that potentially connect the source to the destination. Thus, it might be prohibitively costly to solve the LP $\rm{P1}$ with objective function as in~\eqref{eq:AveCap} over the collection of all paths $\mathcal{P}$. Therefore, we propose a two step procedure: First, we use
a heuristic algorithm, which is a modified version of Dijkstra's  shortest path algorithm and runs in $O\left(N^3\right)$, and efficiently selects  a linear number of paths in $N$ out of a potentially exponential number of paths - we denote this set of  paths as $\mathcal{P}_s$. Then, we solve the LP $\rm{P1}$ with objective function as in~\eqref{eq:AveCap} over $\mathcal{P}_s$ (instead of $\mathcal{P}$) to create the set $\mathcal{P}_k$ in polynomial time in $N$.
The pseudocode of the proposed algorithm can be found in the Appendix, where we let $\V$ be the set of nodes in the network and $P_{\E}$ denote the set of blockage probabilities of the links in $\E$. We next describe at a high level how our algorithm works.

The proposed path selection algorithm is a modified
version of the Dijkstra’s shortest path algorithm that aims to select paths with high average packet rates. Let $p^\star$ be the path with the highest average packet rate among all paths that connect the source to the destination, i.e., $p^\star = \argmax_{p \in \mathcal{P}}C_pS_p$ where
$S_p$ denotes the success (non-failure) probability of path $p$,
\begin{equation}
 S_p = \left( \displaystyle \prod_{(j,i) \in \mathcal{E}_p} \left( 1-p_{j,i} \right ) \right ),
\end{equation}
and $\mathcal{E}_p$ is the set of edges that belong to path $p$.
The path selection procedure is summarized in Algorithm~\ref{alg1} in the Appendix. Algorithm~\ref{alg1} creates the set $\mathcal{P}_s$ by selecting at most $5N$ paths that connect the source to the destination. At every iteration, it calls the function \textit{select\_best\_path} (see Algorithm~\ref{alg2} in the Appendix) to find the path $p^\star$, and adds $p^\star$ to the set $\mathcal{P}_s$. Then, the link with the smallest average packet rate, i.e., the link with the smallest $\ell_{j,i}(1-p_{j,i})$ value on path $p^\star$ is removed from the network so that the algorithm can select a different path in the next iteration. The algorithm continues until either it selects $5N$ paths or there are no remaining paths in the network after link removals\footnote{The reason the algorithm aims to select $5N$ paths is that the optimal average packet rate can always be achieved by activating at most $2N+2$ paths in the network, as proved in Lemma~\ref{lem:LinPaths}. Thus, the algorithm tries to select a sufficient number of paths for the LP $\rm{P1}$ with objective function as in~\eqref{eq:AveCap} such that the set $\mathcal{P}_s$ would include the paths that achieve or approach the optimal average packet rate closely. Although the algorithm might terminate early after link removals, as shown in Example~3, a much smaller than $2N+2$ number of paths might indeed be sufficient to characterize the optimal average packet rate.}. 
As we proved in Lemma~\ref{lem:LinPaths}, the optimal average packet rate can always be achieved by activating at most $2N+2$ paths in the network. Thus, our approach would find the optimal solution, if the set $\mathcal{P}_s$ contains the optimal $2N+2$ paths; yet this is not guaranteed. We construct the set $\mathcal{P}_s$ greedily, by sequentially using the function \textit{select\_best\_path} (see the Appendix) to add ``best paths".  
As we show through the next example, the path $p^\star$ selected by the function \textit{select\_best\_path} is not always the optimal path.

\noindent \textit{Example~5.} Consider the network in Fig.~\ref{permanent_block}. 
There are two paths connecting the source (node $0$) to the destination (node $4$), particularly: $p_1: 0 \rightarrow 1 \rightarrow 3\rightarrow 4$, and $p_2: 0 \rightarrow 2 \rightarrow 3\rightarrow 4$. 
The link capacities are assumed to be $\ell_{1,0} = \ell_{2,0} = 10,~ \ell_{3,2} = 6,~\ell_{3,1} =\ell_{4,3} = 1$.
The link blockage probabilities are assumed to be zero, except the link with capacity $\ell_{3,2}$ for which the blockage probability is $1/2$. If we apply Algorithm~\ref{alg1} (see the Appendix) for this example network, the algorithm selects $\mathcal{P}_s = \{p_2\}$. However, the optimal path from the source to the destination is path $p_1$ with average packet rate of $1$. Algorithm~\ref{alg1} selects $p_2$ because the destination node is the neighbor of node $3$, and the optimal path from the source to node $3$ goes through node $2$. Thus, the algorithm follows the same route to reach the destination.
\hfill $\square$

\begin{remark}
When the link capacities are all equal, the path that has the highest average packet rate is the path with the highest success probability. This problem can be solved efficiently by applying Dijkstra's algorithm and assigning the weight of the edge from node $i$ to node $j$ as $-\log(1-p_{j,i})$ $\forall (j,i) \in [1:N+1]\times [0:N]$. In this case, the function \textit{select\_best\_path} (see the Appendix) is equivalent to Dijkstra's algorithm and finds the optimal path. Moreover, when the link capacities are different and the link blockage probabilities are all equal, the path that has the highest average packet rate is the path with the highest packet rate, i.e., the problem is equivalent to the widest path problem. In this case, Dijkstra's algorithm can be leveraged to efficiently find the optimal path: the link capacities are assigned as the edge weights, and the algorithm uses the bottleneck link capacity instead of the path length. The function \textit{select\_best\_path} is an equivalent algorithm and it indeed finds the optimal path.
\end{remark}
\subsection{Proposed Reinforcement Learning Method}
We here propose an RL method that uses the paths in $\mathcal{P}_k$ obtained from our proposed algorithm in Section~\ref{sec:PathSelection} in order to attain a target packet rate $R^\star$. In particular,
we formulate a single agent RL problem for the scheduling task by defining a Markov decision process (state space, action space and the reward function) as follows:

\noindent$\bullet$ \textit{State Space} $(\mathcal{S})$: Each state vector consists of the packet rates of the selected paths. Therefore, if we denote the rate of the $i$th path in $\mathcal{P}_k$ at step $t$ by $x_{i,t}$, then the state vector at step $t$ is $\mathbf{s}_t = \left[x_{1,t}, x_{2,t},\ldots, x_{k,t}\right]$;

\noindent$\bullet$ \textit{Action Space} $(\mathcal{A})$: Each action vector represents the changes in the packet rates of the selected paths. Formally, the action vector at step $t$ is $\mathbf{a}_t = \left[y_{1,t}, y_{2,t},\ldots, y_{k,t}\right]$, where $y_{i,t}$ denotes the change in the path packet rate of the $i$th path in $\mathcal{P}_k$ at step $t$. Therefore, the state at step $t+1$ is 
    \[
    \mathbf{s}_{t+1} = \mathbf{s}_t+\mathbf{a}_t = \left[x_{1,t}+y_{1,t}, x_{2,t}+y_{2,t},\ldots, x_{k,t}+y_{k,t}\right].
    \]
\noindent$\bullet$ \textit{Reward Function} $(r(\mathbf{s}_t,\mathbf{a}_t))$: We perform reward shaping to encourage the agent to support the desired rate $R^\star$. Towards this end, if the sum of the packet rates through the $k$ paths, denoted by $R$, reaches (or exceeds) $R^\star$, then the agent terminates the episode and receives a reward equal to $1$. Otherwise, the agent receives a reward equal to ${\rm{e}}^R/\kappa$, where $\kappa$ is a large constant value that makes ${\rm{e}}^R/\kappa$ much smaller than $1$.

\noindent \textit{Example~6.} Assume that $k=|\mathcal{P}_k|=2$, where the packet rates of the two selected paths at time $t$ are $1$ and $2$, and the state vector is $\mathbf{s}_t =[1,2]$. For the action vector $\mathbf{a}_t = [0.2,0.7]$, the next state $s_{t+1} = [1.2,2.7]$ represents the updated path packet rates at time $t+1$. \hfill $\square$

\begin{algorithm}
 \caption{Hybrid Scheduling Algorithm.}
 \label{scheduling_alg}
\begin{algorithmic}
\STATE {\bf Initialize:} SAC networks $\theta, \bar{\theta},\phi$
 as in \cite{haarnoja2018soft}.
\STATE {\bf Input:} Desired packet rate $R^\star$ and the set of paths $\mathcal{P}$.\\
\STATE $\bullet$ Create set $\mathcal{P}_s$ as in Algorithm~\ref{alg1}.
\STATE $\bullet$ Create set $\mathcal{P}_k$ by solving the LP $\rm{P1}$ with objective function in~\eqref{eq:AveCap} over $\mathcal{P}_s$.

\FOR{each episode}
    \STATE $\bullet$ Start with zero initial state, i.e., $\mathbf{s}_0 = \mathbf{0}$.
    \FOR{each environment step}
   \STATE $\bullet$ Select an action $\mathbf{a}_t$ from policy $\pi_{\phi}(\mathbf{a}_t\mid \mathbf{s}_t)$ with 1-$\epsilon$ probability. With $\epsilon$ probability, perform informed exploration. If the action makes the next state valid, move to the state $\mathbf{s}_{t+1} = \mathbf{a}_t+\mathbf{s}_{t}$. Otherwise, stay at the current state, i.e., $\mathbf{s}_{t+1} = \mathbf{s}_t$.
        
        \IF{the current packet rate $R$ is greater than or equal to $R^\star$ }
            \STATE $\bullet$ Receive reward $r_t = 1$.
            \STATE $\bullet$ Store the tuple $\left(\mathbf{s}_t,\mathbf{a}_t,r_t,\mathbf{s}_{t+1}\right)$ in the replay buffer, and then terminate the episode.
        \ELSE
            \STATE $\bullet$ Receive reward $r_t = {\rm{e}}^{R}/\kappa$ and store the tuple $\left(\mathbf{s}_t,\mathbf{a}_t,r_t,\mathbf{s}_{t+1}\right)$ in the replay buffer.
        \ENDIF

    \ENDFOR
    \FOR{each gradient step}
    \STATE $\bullet$ Update the network parameters using gradient descent.
    \ENDFOR

\ENDFOR

\end{algorithmic}
\end{algorithm}

We note that the definitions of state space, action space and reward function do not assume  knowledge of the link capacities. We consider an episodic RL method for this continuous control problem, where each episode lasts for a finite horizon $T$. However, if the agent reaches the specified desired packet rate $R^\star$ during an episode, then the episode ends early. Moreover, the next state has to be a physically feasible state, i.e., the fraction of time each path is used should satisfy the constraints of the LP $\rm{P1}$ in~\eqref{capacity_paths}. This means that all path packet rates have to be non-negative, and that a node cannot transmit or receive more than $100\%$ of the time. Thus, if the action vector makes the next state invalid, it is assumed that the agent stays at the current state. We assume that the environment can determine whether a state is valid or not since if the path packet rates at a particular state are invalid, then the network cannot support them and it enters outage (e.g., it drops packets due to queue congestion).
In a real deployment, the source node can be the agent, and it can acquire the packet rates of the selected paths through TCP feedback and by observing packet drops. It can then accordingly adjust the path packet rates at each step, and if it reaches the desired packet rate $R^\star$, it can terminate the episode.
In summary, the agent updates the packet rates of the selected paths at each step through the action vector, unless the action vector makes the next state invalid. In the latter case, the path packet rates are not changed and the agent stays at the current state. 

Our goal is to design an algorithm to provide resilience against network disruptions.
Thus, we propose an {\em informed exploration} technique such that the agent can explore the space in a more informed way compared to a random exploration. In particular, the agent needs to explore unblocked paths effectively without using any side information - understanding if there are blockages in the network and bypassing the blocked paths is part of its learning process. Thus, the agent selects the action vector from its policy $\pi$ with $(1-\epsilon)$ probability, and with $\epsilon$ probability the agent performs exploration. At the beginning of the experiment, the agent sorts the paths in $\mathcal{P}_k$ according to their success (non-failure) probability. At every exploration step $t$, the agent selects a random number $m_t$ of paths that have the highest success probability where $1\leq m_t \leq k$. It then assigns the action values for the selected $m_t$ paths by sampling from a uniform distribution between $0$ and $1$. The action values for the remaining paths are zero. For example, if the agent decides to explore at step $t$, it randomly selects a value $m_t$, e.g., $m_t=3$. Then, the agent selects $3$ paths that have the highest success probability and sets the action vector by following the above procedure. This exploration method might help the agent to detect and explore the unblocked paths. 
The proposed scheduling algorithm is provided in Algorithm \ref{scheduling_alg}.

\section{Performance Evaluation}\label{section_evaluation}
In this section, we numerically evaluate the proposed scheduling algorithm and compare it against alternative algorithms with
respect to different performance metrics, as we discuss next.
\subsection{Experiment Settings}
\noindent\textbf{Simulated Networks.} We used the same neural network architecture and the hyperparameters in~\cite{haarnoja2018soft} for the SAC algorithm. We list the hyperparameters in Table~\ref{parameters} and the source code of our implementation is available online\footnote{\nolinkurl{github.com/minedgan/hybrid_mmwave_scheduling}}.
\begin{table}
\caption{SAC Hyperparameters.\label{parameters}
}
\begin{center}
\begin{tabular}{|p{7cm}||p{1.5cm}|}
\hline
  \textbf{Parameter}
  & \textbf{Value}
 \\
\hline
 \hline
 Optimizer & Adam \\
 \hline
 Learning rate & $3\cdot 10^{-4}$  \\
 \hline
 Discount & $1$  \\
 \hline
 Replay buffer size & $10^6$  \\
 \hline
 Number of hidden layers (all networks) & 2\\
 \hline
 Number of hidden units per layer & 256 \\
 \hline
 Number of samples per minibatch & 32\\
 \hline
 Nonlinearity & ReLU\\
 \hline
 Target smoothing coefficient & 0.005\\
\hline
\end{tabular}
\end{center}
\end{table}
In the experiments, we considered a mmWave network with $N=25$ relay nodes and at least $1,000$ paths were randomly generated between the source node and the destination node. The coordinates of each node were sampled from a uniform distribution between $0$ and $100$, and the capacity of each link was generated from the Rician distribution based on the omni-directional path loss~\cite{ThomasVTC14}. 

We trained the agent for $200$ episodes and each episode had a time horizon of $T = 500$, i.e., each episode lasted at most $500$ time steps. In our evaluations, we considered two cases: (i) blockage in the static network; and (ii) blockage in the time-varying network. In both cases, we assigned blockage probability to each link according to~\eqref{block_prob}. We generated each $\lambda_{j,i}$ in~\eqref{block_prob} by sampling from a uniform distribution between $200$ and $600$ so that a high fraction of paths could be blocked. During the training, at every $10$ episodes, a new set of links was blocked based on the assigned blockage probabilities, and these links remained blocked until a new set was selected. In the static network, the link capacities stayed constant throughout the training (except for the capacity of the blocked links). The average link capacity was $7.22$, and the approximate capacity of the initial network was equal to $9.52$ as computed by solving LP $\rm{P1}$ in~\eqref{capacity_paths}. In the time-varying case, at every episode, each link capacity was sampled from a Gaussian distribution with variance equal to $1$, and mean value equal to the capacity of that link in the static network. The variation in the link capacities captures the nodes mobility or the varying channel conditions. In both cases, the desired rate $R^\star$ was set to $70\%$ of the network approximate capacity $\widebar{\mathsf{C}}$. As the link capacities changed, the network capacity and $R^\star$ also changed. We note that, although in our experiments we calculated $\widebar{\mathsf{C}}$ to display the performance, this is not  needed in a real deployment: as our evaluation shows, if a desired rate is achievable, a good agent achieves it.

\noindent\textbf{Path Selection.} We used the following procedure to create $\mathcal{P}_k$. In the static case, we created different sets of $\lambda_{j,i}$ values which result in different blockage probabilities. For each set, we applied Algorithm~\ref{alg1} and then, solved LP $\rm{P1}$ in~\eqref{capacity_paths} with objective function as in~\eqref{eq:AveCap}. We then took the union of the paths selected by the optimal solution for each set of $\lambda_{j,i}$ values to create $\mathcal{P}_k$. In the time-varying case, we created different sets of $\lambda_{j,i}$ values (different blockage probabilities) and link capacities where each link capacity was sampled from a Gaussian distribution with variance equal to $1$, and mean value equal to the capacity of that link in the static network. For each set, we again applied Algorithm~\ref{alg1} and then, solved LP $\rm{P1}$ in~\eqref{capacity_paths} with objective function as in~\eqref{eq:AveCap}. We then took the union of the paths selected by the optimal solution for each set.
For both the static and the time-varying cases, we ensured that at least $10$ paths were included in $\mathcal{P}_k$. The reason we performed this procedure is to ensure that there  will be a sufficient number of paths that the RL agent can exploit. Moreover, this procedure might improve the adaptability of the scheduling algorithm to different blockage realizations. The agent adjusts the packet rates of the paths in $\mathcal{P}_k$, and at a specific time, the network rate is calculated as the sum of the packet rates of the paths in $\mathcal{P}_k$.  
In the time-varying case, since the link capacities might drastically change throughout the training, we occasionally applied the aforementioned path selection procedure to select a new set of paths for the agent. Towards this end, the agent uses the received rewards as a signaling method for the path selection. Particularly, if the agent does not receive reward $1$ for a certain number of time steps $t_s$, which indicates that the agent does not exceed the desired packet rate $R^\star$, then the agent selects a new set of paths by following the aforementioned path selection procedure. In our evaluations, we selected $t_s=5T$ ($5$ episodes).

During training, at each time step, an action vector was sampled from the policy which had a Gaussian distribution in our evaluations. After sampling the action vector, we applied the hyperbolic tangent function (tanh) to the sample in order to bound the actions to a finite interval~\cite{haarnoja2018soft}. Moreover, we performed action clipping such that if the elements of the action vector (after applying tanh function) were less than $10^{-3}$, these elements were taken as $0$'s. The action clipping is necessary because the agent might need to take zero actions for some paths, particularly if these paths are blocked, and it is not possible to instantiate zero action by sampling from a continuous distribution without action clipping. For the exploration, we chose $\epsilon = 0.01$ and for the reward function, we chose $\kappa = 10^7$ (see also Algorithm~\ref{scheduling_alg}).

\subsection{Performance Metrics}
We evaluated the performance of the proposed algorithm by using the following two metrics.

\noindent $\bullet\ ${\em Average Training Rate.}  This is the average packet rate achieved during training, which has important practical implications: it captures whether and how fast the network is able to support reasonable packet rates while training; it thus indicates whether it is possible to perform training online, while still utilizing the network. Towards this end, we trained five different instances of the algorithm with different random seeds. At each instance, we generated a different set of link capacities and blockage probabilities, as well as a different topology where at least $1,000$ paths were randomly generated. For each instance, we examined the average packet rate achieved in every episode during training. We found the average packet rate achieved in an episode by taking the average of the packet rates achieved at each time step during that episode. For episodes terminating earlier than the time horizon, the last rate was assumed to be maintained for the rest of the episode.
We finally took the average of the packet rates over these five instances.

\noindent $\bullet\ ${\em Evaluation Rate.} At the end of each training episode, we performed evaluation and found the packet rate achieved by the agent. Thus, this process can be considered as the validation of the policy during training. While finding the evaluation rate at the end of an episode, the agent started from a zero initial state and adjusted the path packet rates by using its current policy. We again used $T=500$ as time horizon: if the agent exceeded the desired packet rate during the evaluation, it stopped; otherwise, the final packet rate was the rate achieved at the last time step $t=500$. The agent did not perform informed exploration during this process. We note that the policy of the agent in the SAC algorithm is stochastic, hence the agent can take different actions at the same state. Thus, we repeated the same evaluation procedure for five times and took the average of the final packet rates to find the evaluation rate at that episode.

\subsection{Alternative Methods} 
\label{sec:Baseline}
Although various routing algorithms exist in the literature, we cannot compare our proposed method with them: these existing algorithms are tailored to general networks and do not consider the scheduling constraints in mmWave networks, or they rely on centralized knowledge of the link capacities. Therefore, we compared our proposed method with the following algorithms.
\\
$\bullet$  \textbf{Baseline 1}: 
In this baseline approach, we evaluate the performance of the centralized approach by solving LP $\rm{P1}$ in~\eqref{capacity_paths} and leveraging the paths that are activated in the optimal solution and their corresponding schedule to support the desired rate. We select paths and find their optimal schedule only at the beginning and use them throughout the experiment. Although this baseline algorithm distributes the traffic across multiple paths both over space and time, it does not consider resilience against link failures.
 \\
$\bullet$  \textbf{Baseline 2}: 
In this approach, we first solve LP $\rm{P1}$ in~\eqref{capacity_paths} and find the network approximate capacity $\widebar{\mathsf{C}}$. We then turn LP $\rm{P1}$ in~\eqref{capacity_paths} into a feasibility problem where the objective function is set to $0$ and an additional constraint is added, namely
\begin{equation*}
    \displaystyle\sum_{p \in \mathcal{P}} x_p \mathsf{C}_p = R^\star,
\end{equation*}
where $R^\star = 0.7 \widebar{\mathsf{C}}$.
Thus, this constraint ensures that the achieved packet rate is equal to the desired rate. We leverage the paths that are activated in the optimal solution as well as their corresponding schedule to support the desired rate. In this baseline approach, we again select paths and find the corresponding schedule only at the beginning and leverage them throughout the experiment. This baseline approach evaluates the performance of the centralized approach when it tries to achieve only the desired rate instead of the approximate capacity. 

We highlight that the paths selected by the above baseline approaches might be different since the first approach maximizes the rate while the second one is trying to achieve the desired rate.
\\
$\bullet$ \textbf{Highest Capacity (HC): } At the beginning of each training instance, we solve LP $\rm{P1}$ in~\eqref{capacity_paths} and select the $k$ paths with the highest activation time among the paths activated by the optimal solution. We leverage these paths while training the RL agent in that training instance. We note that, even though the selected paths have high packet rates, this path selection procedure does not consider resilience against link failures. Our aim here is to understand if simply relying on high-capacity paths provides packet rate guarantees.
\\
$\bullet$\textbf{Random Selection (RS): } At each training instance, we randomly select $k$ paths out of all paths and leverage them while training the RL agent in that training instance.

\subsection{Numerical Evaluation}
We here compare, through simulation results, the performance of our proposed algorithm
versus the algorithms in Section~\ref{sec:Baseline}. We observe the following results.\\
$\bullet$\textbf{Blockage in a static network. } We first compare the performance of our scheduling algorithm against Baseline 1 and Baseline 2. As shown in Fig.~\ref{static}, both baseline methods could not support the desired rate because they select the paths and their schedule without considering resilience against link failures.
\begin{figure*}
     \centering
     \begin{subfigure}[b]{0.3\textwidth}
         \centering
         \includegraphics[width=\textwidth]{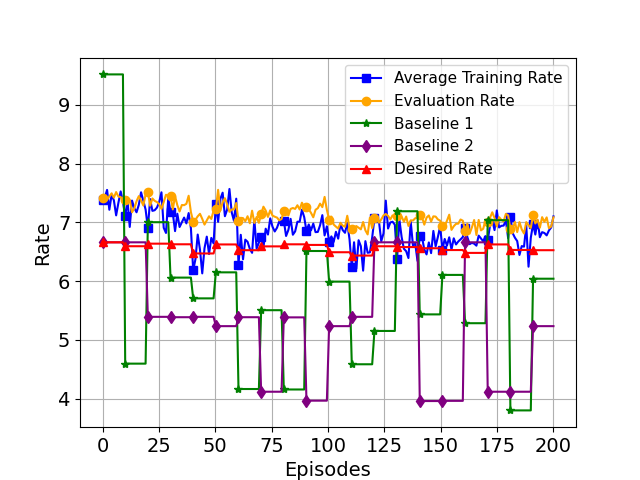}
         \caption{}
         \label{static}
     \end{subfigure}
     \begin{subfigure}[b]{0.3\textwidth}
         \centering
         \includegraphics[width=\textwidth]{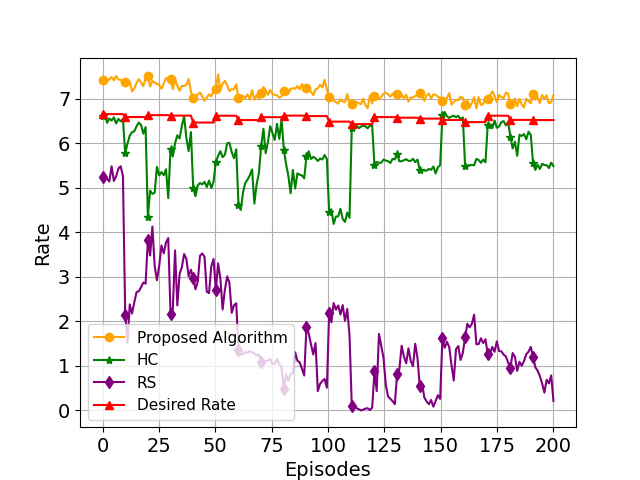}
         \caption{}
         \label{static_path_selection}
     \end{subfigure}
    \begin{subfigure}[b]{0.3\textwidth}
         \centering
         \includegraphics[width=\textwidth]{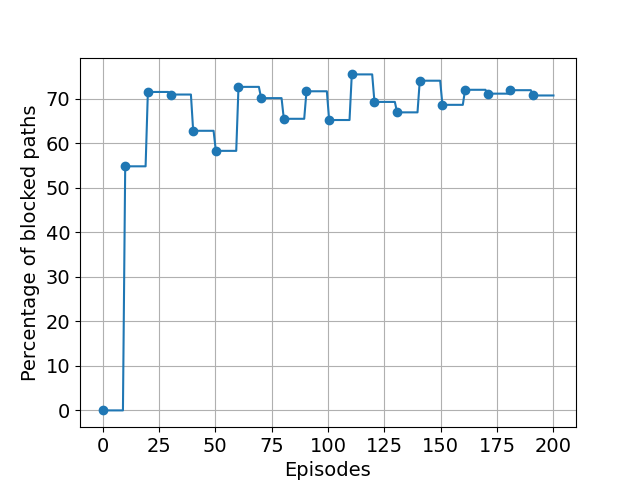}
         \caption{}
         \label{perc_blocked_paths}
     \end{subfigure}
        \caption{Performance of the proposed algorithm and alternative algorithms (static network).}
        \label{comparison}
        \vspace{-0.1in}
\end{figure*}
When the paths get blocked, the allocated time slots of those paths remain idle, which results in wasted resources. However, our proposed scheduling algorithm selects proactively resilient paths with high packet rates, and adjusts the path packet rates to effectively exploit the unblocked paths. In Fig.~\ref{perc_blocked_paths}, the average percentage of number of blocked paths is shown and we can observe that a high percentage of paths is blocked on average over $5$ training instances.
We highlight that the agent does not use any side information - understanding if there are blockages in the network and bypassing the blocked paths is part of its learning process.

We also compared the performance of our path selection algorithm against the HC and RS methods. As shown in Fig.~\ref{static_path_selection}, both these methods could not support the desired rate. Indeed, RS randomly selects paths without considering their capacity or failure probability. Although HC selects paths that have high capacity, it does not consider resilience against link failures. If the selected paths have high failure probabilities, the RL agent might need to operate a substantially small number of paths in $\mathcal{P}_k$, which might prevent it from supporting the desired rate if the available paths are not strong enough. Differently, the paths selected by our algorithm allow the agent to support the desired rate because they have high packet rates and low failure probability.
\\
$\bullet$\textbf{Blockage in a time-varying network. } We first compare the performance of our scheduling algorithm against Baseline 1 and Baseline 2. The profile of the blocked paths is the same as in the static case shown in Fig.~\ref{perc_blocked_paths}. As shown in Fig.~\ref{time_varying}, both baseline methods could not support the desired rate for the same reasons as in the static case.
\begin{figure*}
     \centering
     \begin{subfigure}[b]{0.3\textwidth}
         \centering
         \includegraphics[width=\textwidth]{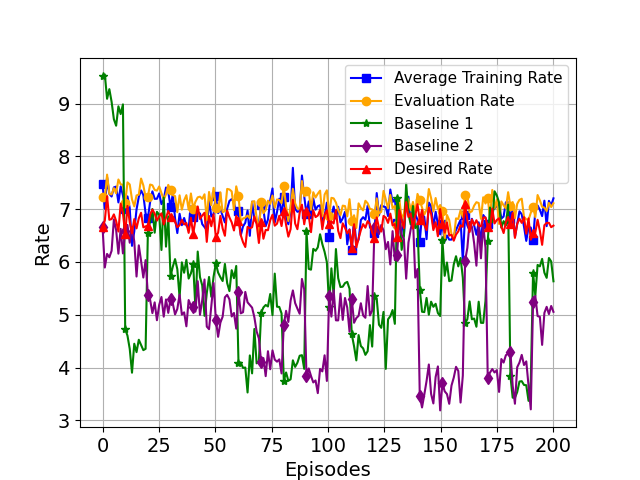}
         \caption{}
         \label{time_varying}
     \end{subfigure}
     \begin{subfigure}[b]{0.3\textwidth}
         \centering
         \includegraphics[width=\textwidth]{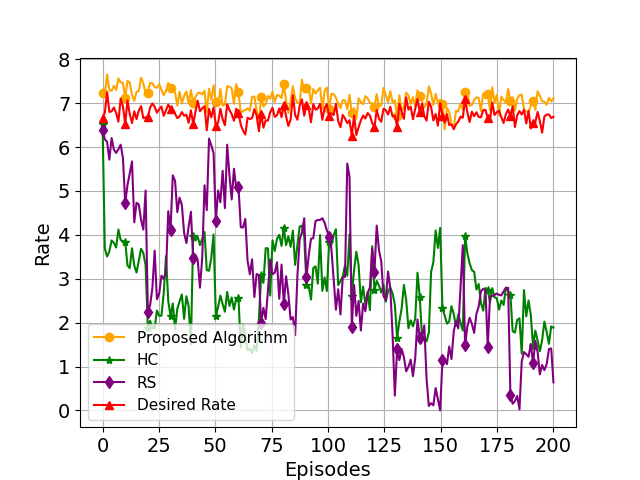}
         \caption{}
         \label{timevarying_path_selection}
     \end{subfigure}
        \caption{Performance of the proposed algorithm and alternative algorithms (time-varying network).}
        \label{comparison2}
        \vspace{-0.1in}
\end{figure*}
Moreover, although the selected paths have high capacity initially, they might not remain strong enough to support the desired rate as the link capacities vary in later episodes.  However, our proposed scheduling algorithm adjusts the path packet rates at every time step to find a good schedule. It adapts to blockage and channel variations by effectively exploiting the unblocked paths. One advantage of our algorithm is that we occasionally apply the path selection procedure to select a new set of paths for the
RL agent as the link capacities might drastically
change throughout the experiment. However, the agent uses the received rewards as a signaling method for the path selection instead of retrieving the channel and topology information frequently, which might be substantially costly in volatile environments. Since the baseline methods do not have a similar signaling method, they need to retrieve the channel and topology information frequently to determine when to select a new set of paths. 
Even though we propose the aforementioned path selection algorithm for varying channel conditions, the agent did not select a new set of paths over five instances (total of $1,000$ episodes) in our experiments. This shows that the proposed algorithm robustly adapts to varying channel conditions and network disruptions by leveraging the initially selected paths.

We also compared the performance of our path selection algorithm against the HC and RS methods. As shown in Fig.~\ref{timevarying_path_selection}, both these methods could not support the desired rate for the same reasons as in the static case.
Since our algorithm can occasionally apply the path selection procedure to select a new set of paths as the link capacities change, it ensures that the agent has strong and resilient paths to operate.


\appendix
\section{}\label{appendix_edgedis}
We start by considering the optimal \mbox{worst-case} packet rate.
We let $m$ denote the capacity of the links. 
We find an optimal solution for the LP $\rm{P1}$ with the objective function in~\eqref{eq:WSCap} for the input network. 
We denote such an optimal solution with $(t^\star,x^\star)$, where $t^\star$ is the optimal \mbox{worst-case} packet rate and $x^\star$ denotes an optimal activation time for the paths in the network.
If the solution does not include any overlapping paths, i.e., all of the activated paths are edge-disjoint, then we found a solution that satisfies (P1) in Theorem~\ref{results}. If the solution includes overlapping paths, then we create another solution in which all of the activated paths are edge-disjoint. The procedure is summarized in Algorithm~\ref{L2_alg}.
\begin{algorithm}
 \caption{Edge-disjoint Paths Solution for the Optimal Worst-case Packet Rate.}
 \label{L2_alg}
\begin{algorithmic}
\STATE \textbf{Input:} An $N$-relay arbitrary Gaussian 1-2-1 network, the set of links $\mathcal{E}$, and $k_{\ell}$. 
\STATE {\bf Output:} A set of edge-disjoint paths $\mathcal{S}'$ and the corresponding path activation times $x$.
\STATE Solve LP $\rm{P1}$ with the objective function as in~\eqref{eq:WSCap} for the input network and let $x^\star$ denote the optimal path activation times and $t^\star$ denote the optimal \mbox{worst-case} packet rate. 
\STATE $\mathcal{S}$ denotes the set of operating paths, i.e., $\mathcal{S} = \{p \in \mathcal{P}: x^\star_p >0\}$, $\mathcal{S}'\leftarrow \varnothing$, and $\mathcal{H}\leftarrow \varnothing$.
\IF {$\mathcal{S}$ only includes edge-disjoint paths}
    \STATE Return $\mathcal{S}$ and $x^\star$.
\ELSE
    \STATE Identify the set of edge-disjoint paths $\mathcal{U} = \{\alpha_1,\dots,\alpha_r\}$ in $\mathcal{S}$, and then $\mathcal{S}' \leftarrow \mathcal{S}' \cup \mathcal{U}$.
    \STATE For each link $e \in \mathcal{E}$, find the set of paths $\mathcal{D}_e$ passing through that link. 
    \STATE For each link $e \in \mathcal{E}$ with $|\mathcal{D}_e| > 1$, choose a path $p$ from $\mathcal{D}_e$ and add it to the set $\mathcal{H}$ such that all paths in $\mathcal{H}$ are edge-disjoint and it has the highest number of edge-disjoint paths.
    \STATE  $\mathcal{S}' \leftarrow \mathcal{S}' \cup \mathcal{H}$, and then
     assign activation time $x_p = t^\star/m(|\mathcal{S}'|-k_{\ell})$ to each path $p \in \mathcal{S}'$. The remaining path activation times are zero, i.e., $x_p = 0$ $\forall p \notin \mathcal{S}'$. 
    \STATE Return $\mathcal{S}'$ and $x$.
\ENDIF
\end{algorithmic}
\end{algorithm}
We now show that the solution that Algorithm~\ref{L2_alg} finds is a feasible solution for LP $\rm{P1}$ with the objective function in~\eqref{eq:WSCap}, and its \mbox{worst-case} rate is $t^\star$. 
We first note that $|\mathcal{S}'|-k_{\ell} >0$ by construction, otherwise, the \mbox{worst-case} rate would be zero in the original solution.
By construction, all of the paths in $\mathcal{S}'$ are edge-disjoint. When $k_{\ell}$ links are blocked, each link must be from a different path in the \mbox{worst-case} scenario. Thus, $k_{\ell}$ paths will be blocked in the worst-case (all paths have the same capacity, thus which set of $k_{\ell}$ paths is blocked does not affect the \mbox{worst-case} result). The packet rate achieved by the remaining $(|\mathcal{S}'|-k_{\ell})$ paths is
\begin{equation}\label{sol_worst}
x_p m (|\mathcal{S}'| - k_{\ell}) = \frac{t^\star(|\mathcal{S}'|-k_{\ell})m}{m(|\mathcal{S}'|-k_{\ell})} = t^\star,
\end{equation}
which is indeed the optimal \mbox{worst-case} packet rate.
We next show that all of the constraints in LP $\rm{P1}$ are satisfied with the construction in Algorithm~\ref{L2_alg}. 
By construction, we have that $x >0$, thus the constraints in $\rm{(P1}a)$ are satisfied. For the constraints in $\rm{(P1}b)$ and $\rm{(P1}c)$, we note that $f^p_{p.nx(i),i} = 1$ $\forall i \in [0:N]$ and $f^p_{i,p.pr(i)} = 1$ $\forall i \in [1:N+1]$ because we consider the equal link capacity case. Therefore, we here show that $x$ satisfies the constraints in $\rm{(P1}b)$ and $\rm{(P1}c)$ for the source and the destination nodes, particularly we show that $\sum_{p \in \mathcal{P}} x_p \leq 1$.
If there is no blockage in the network, the packet rate from the source to the destination is $m$. 
We now note that we have $|\mathcal{S}'|$ ``clusters" in the network. A cluster might be a single edge-disjoint path from $\mathcal{U}$ or the set of paths that share an edge with a path $p \in \mathcal{H}$. In the \mbox{worst-case} scenario, every blocked link blocks one of the clusters in the network and decreases the packet rate by $m/|\mathcal{S}'|$. Thus, for $k_{\ell}$ link blockages, we have $|\mathcal{S}'|-k_{\ell}$ clusters in the worst-case such that each cluster receives a rate of $m/|\mathcal{S}'|$ from the source, thus the \mbox{worst-case} packet rate is $t^\star = m(|\mathcal{S}'|-k_{\ell})/|\mathcal{S}'|$. By using this property, we can show that the $x_p$'s output by Algorithm~\ref{L2_alg} satisfy the constraints in $\rm{(P1}b)$ and $\rm{(P1}c)$ at the source and the destination as follows,
\begin{equation*}
    \sum_{p \in \mathcal{P}} x_p= \frac{t^\star |\mathcal{S}'|}{m(|\mathcal{S}'|-k_{\ell})} 
    \implies \sum_{p \in \mathcal{P}} x_p= \frac{m(|\mathcal{S}'|-k_{\ell})}{|\mathcal{S}'|}\frac{|\mathcal{S}'|}{m(|\mathcal{S}'|-k_{\ell})} = 1.
\end{equation*}
Since the constraints are satisfied at the source and the destination, then the constraints for all of the other nodes in $\rm{(P1}b)$ and $\rm{(P1}c)$ are also satisfied (because of the assumption of equal link capacities). Therefore, $x$ from Algorithm~\ref{L2_alg} gives a feasible solution for LP $\rm{P1}$ with the objective function in~\eqref{eq:WSCap}. 
This concludes the proof of (P1) in Theorem~\ref{results} for the \mbox{worst-case} packet rate. 

We now prove (P1) in Theorem~\ref{results} for the optimal average packet rate of a network with equal link capacities. We can rewrite the objective function in~\eqref{eq:AveCap} as $\sum_{p \in \mathcal{P}}b_px_p,$
where $b_p$'s are the constants that depend on the blockage probabilities and on the path capacities. Due to the constraints at the source and the destination in $(\rm{P1}b)$ and $(\rm{P1}c)$, we have that $\sum_{p \in \mathcal{P}}x_p \leq 1.$
Therefore, the objective function can be maximized by setting $x_{p'} = 1$ for $p' = \arg\max_{p \in \mathcal{P}} b_p$, and  $x_p = 0$ $\forall p\in \mathcal{P}\backslash p'$\footnote{If there are multiple paths whose $b_p$'s are maximum, we can choose one of them arbitrarily and set its activation time to $1$.}. Since we can always activate a single path to maximize the average rate, this concludes the proof of Theorem~\ref{results} (P1).

\begin{algorithm}
\caption{Path Selection Algorithm.}\label{alg1}
\begin{algorithmic}
\STATE $\mathcal{P}_s \leftarrow \varnothing$.
\WHILE{$|\mathcal{P}_s| < 5N$}
\STATE weight, previous $ = \textit{select\_best\_path}(\V,\E, P_{\mathcal{E}})$ .
\STATE $p^\star \leftarrow$ empty sequence and $u \leftarrow N+1$. 
\WHILE{$u$ is DEFINED}
\IF{previous[$u$] = UNDEFINED and $u \neq 0$}
    \STATE Return $\mathcal{P}_s$.
\ENDIF
\STATE Add $u$ to the beginning of $p^\star$, and then $u \leftarrow \text{previous[}u\text{]}$.
\ENDWHILE
\STATE $\mathcal{P}_s \leftarrow \mathcal{P}_s \cup p^\star$, then find the link with the smallest $\ell_{j,i}(1-p_{j,i})$ value on $p^\star$, and remove that link from the network.
\ENDWHILE
\end{algorithmic}
\end{algorithm}

\begin{algorithm}
\caption{Selection of the Path with the Highest Average Packet Rate.}\label{alg2}
\begin{algorithmic}
\STATE \textbf{function} \textit{select\_best\_path}$(\V,\E, P_{\mathcal{E}}):$
\STATE $\K \leftarrow \varnothing$.
\FOR{each vertex $v$ in $\V$}
\STATE weight[$v$] = $-\infty$, capacity[$v$] = $-\infty$, success[$v$] = $1$, and previous[$v$] = UNDEFINED.
\ENDFOR
\STATE weight[$0$] = $\infty$, capacity[$0$] = $\infty$, and success[$0$] = $1$.
\WHILE{$\K \neq \V$}
\STATE Pick $u \notin \K$ with the largest weight, and then $\K \leftarrow \K \cup u$.
\FOR{each neighbor $v$ of $u$ such that $v \notin \K$}
\STATE $a \leftarrow$ success[$u$]*$(1-p_{v,u})$, $b \leftarrow \min(\text{capacity[$u$]},\ell_{v,u})$, and $c \leftarrow \max(\text{weight[$v$]},ab)$.
\IF{$c > \text{weight[$v$]}$ and weight[$u$] $\neq -\infty$ }
\STATE weight[$v$] $\leftarrow c$, success[$v$] $\leftarrow a$, capacity[$v$] $\leftarrow b$, and previous[$v$] $\leftarrow u$.
\ENDIF
\ENDFOR
\ENDWHILE
\STATE Return weight, previous.
\end{algorithmic}
\end{algorithm}


 




\bibliographystyle{IEEEtran}
\bibliography{IEEEabrv,bibliography}

\vfill
\end{NoHyper}
\end{document}